\documentclass[aps,prl,twocolumn,showpacs,groupedaddress,amsmath,amssymb]{revtex4}

\usepackage{graphicx}

\def \be{\begin{displaymath}}
\def \ee{\end{displaymath}}              

\def \ben{ \begin{equation} }
\def \een{ \end{equation}   }            

\def \bea{\begin{eqnarray*}}             
\def \eea{\end{eqnarray*}}

\def \bean{\begin{eqnarray}}             
\def \eean{\end{eqnarray}}

\def \cC{ {\cal C} }

\def \cR{ {\cal R} }
\def \cH{ {\cal H} }
\def \cO{ {\cal O} }
\def \cZ{ {\mathbf{Z}} }

\def \Ref#1{(\ref{#1})}
\def \eref#1{Eq.\ \Ref{#1}  }

\def \eps{\varepsilon}

\def \e{ {\rm e}}

\def \br{ {\mathbf r} }
\def \bk{ {\mathbf k} }

\def \inv{ ^{-1} }
\def \dag{^\dagger}

\def \invb#1 { \frac{1}{#1} }

\def \ket#1{ {\left| #1 \right\rangle} }

\def \sbrackets#1{ \langle #1 \rangle }
\def \sket#1{ | #1 \rangle }
\def \lket#1{ | #1 \rangle_L }
\def \sbra#1{ {\langle #1 |} }

\def \hw#1{ { | #1 | } }    

\def \fr#1#2{ \frac{#1}{#2} }
\def \binomial#1#2{ \left( {#1} \atop {#2} \right) }

\begin{document}
\bibliographystyle{prsty}
\input epsf
\title{Quantum Error Correction in Correlated Quantum Noise} 
\author{Rochus Klesse and  Sandra Frank}
\affiliation{ Universit\"at zu K\"oln, Institut f\"ur Theoretische Physik,
     Z\"ulpicher Str. 77, D-50937 K\"oln, Germany}
\date{May 20, 2005}
\begin{abstract}
We consider quantum error correction of quantum noise that is created by
a local interaction of qubits with a common bosonic bath. The possible
exchange of bath bosons between qubits gives rise to spatial and temporal
correlations in the noise.
We find that these kind of noise correlations have a strong 
negative impact on quantum error-correction. 
\end{abstract}
\maketitle

The superiority of quantum computation over conventional computation
relies on the fact that a quantum-bit (qubit) register can be in the
superposition of a very large number of classical computational
states. At the same time, maintaining coherence of this highly superpositional 
state poses also the main obstacle for the realization of a quantum computer.  

For a small number of qubits this difficulty can be overcome by simply
reducing the coupling to environmental degrees of freedom, as has been
demonstrated by several groups for different physical implementations.
With an increasing number of qubits, it will, however, become 
extremely difficult to reach the required coherence in that way \cite{unruh}. 
It is therefore common opinion that a scalable implementation of a quantum
computer must use some error correction scheme that recovers the
quantum state after it has been distorted by external noise. 

The existence of error correcting schemes for quantum states,
which was shown independently by Shor \cite{shor} and Steane
\cite{steanePRL}, is a remarkable fact and has been crucial for the
development of the field.
The key ideas presented in their work rapidly evolved to a beautiful
theory of quantum error correcting codes and subsequently to the
concept of fault tolerant quantum computation \cite{qec_general}. 

Quantum error correcting schemes are usually designed for the 
independent error model, which by definition does not exhibit
correlations between noise of different times and locations.
From a physical point of view, this requirement
is rather annoying, since in general qubits do interact with a {\em
common} environment which necessarily introduces some amount of
correlations in the noise.   
To be more specific, in many if not all situations the qubits weakly
interact with a common thermal bath of extended bosons (photons
and/or phonons). The exchange of bosons between qubits will then
cause spatial and temporal error correlations that violate the
condition of error independence. Indeed, it has been shown
\cite{palma} that these kind of processes can lead to drastically 
enlarged or reduced decoherence of certain states. 

To which extent do such error correlations interfere with 
quantum error correction? We have analyzed this problem for
optimal Calderbank-Shor-Steane (CSS) quantum error-correcting codes 
of variable length $n$ (number of physical qubits) and size $k$ (number
of logical qubits).
As physical noise-model, we use a reduced spin-boson model
consisting of $n$ spins -- describing an $n$-qubit register -- coupled
to a common bosonic bath \cite{palma,reina}. The amount of noise correlations   
can be controlled by the inter-spin distance $r$. 
Within this framework, we study how code states transform during
spin-boson interaction and a subsequent error-correcting
operation. The distance between the resulting code state and the
initial one -- in the sequel denoted as residual error $\Delta$
[cf.~Eq.~\Ref{delta_definition}] --
serves as  a measure for the error-correcting performance of the code. 

\begin{figure}
  \begin{center}
    \epsfxsize=7.0cm
    \epsffile{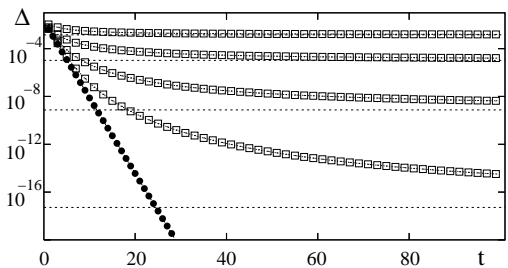}
     \vspace{-0.3cm}
    \caption{Residual error $[\Delta]_t^n$ of $[n,k,d=2t+1]$ CSS codes
    as function of $t$ for $\Gamma_0=0.01$, $k/n = 0.062$, and $t/n = 0.05$.
    The values of $\Gamma_r$ are $.01$, $.005$, $.0025$, $.00125$
    (squares, top to down), and $\Gamma_r=0$ (dots). The 
    dashed lines are the asymptotic values of the residual errors.
    \label{fig1}}
  \end{center}
  \vspace{-0.5cm}
\end{figure}

Our main finding is that quantum error correction with CSS
codes is substantially hampered by the kind of noise-correlations
captured in our model. This becomes evident by the fact that 
for any fixed information rate $k/n>0$ the residual error
$\Delta$ approaches a finite constant in the limit $n\to\infty$, unless the
spin-boson coupling strictly vanishes or $r$ is infinite
(cf.~Fig.~\ref{fig1}).   
Using a simple scaling argument we conclude that for a wide range of
model parameters CSS codes cannot provide the accuracy
needed for large scale quantum computations. 

In contrast to related studies \cite{duan_guo}, here the spin-boson
coupling is treated in a non-perturbative manner, which we find to
be indispensable in the large $n$ limit. Substantial progress towards
error correction beyond the independent error model has been made in
very recent work \cite{terhal_burkard,aliferis}. We will briefly
comment our results in light of this new work at the end of this
Letter. An extended discussion will follow in a future publication
\cite{to_be}. 

We begin with the physical model for the $n$-qubit register. It is
defined by the Hamiltonian 
\vskip-0.55cm
\be
H = \sum_{l=0}^{n-1} \fr{\epsilon}{2} \sigma_{z,l}\: + \: \sum_{\bk}
\omega_\bk b\dag_{\bk} b_{\bk}  \: + \: \sum_{\bk,l}
\sigma_{z,l} ( g_{|\bk|} \e^{i \bk \cdot \br_l} b\dag_{\bk} \:+ \:\mbox{H.c.}\:)
\ee 
\vskip-0.2cm
\noindent of $n$ spin-$\fr{1}{2}$ particles (qubits) at positions $\br_l$ interacting
with three-dimensional bosonic modes with creation (annihilation) operators
$b_\bk\dag$ ($b_\bk$) and energies $\omega_\bk = |\bk|$ \cite{units}.
$\sigma_{z,l}$ denotes the Pauli spin $\sigma_z$ operator acting on the $l$th spin,
and $\epsilon$ is the Zeeman energy. The spin-boson couplings $g_{|\bk|}$
may be characterized  as usual by a spectral function
$
J(\omega) := \sum_\bk \delta(\omega_\bk - \omega) |g_{|\bk|}|^2 \equiv A \omega^s
\e^{-\omega/\Omega}
$,
with a cut-off frequency $\Omega$, a constant $A$ of
appropriate dimension, and a positive parameter $s$ \cite{books}. 
Note that the spin-boson couplings do not lead to energy dissipation
but to dephasing of the spin system. 

The spin-boson model given by Hamiltonian $H$ is quite suitable for
our purposes because it shows full decoherence and is still
analytically solvable \cite{palma,reina}.  
We assume that initially the system is in a product state $\rho(0)\otimes \rho_b$
of a spin state $\rho(0)$ and a thermal state $\rho_b$ of the bosonic
bath at temperature $T$. Following Refs.~\cite{palma,reina} we
determine the reduced density matrix $\rho(\tau)$ of the
spin system at some subsequent time $\tau$ to be  \cite{to_be}
\ben\label{projector_rho}
\rho(\tau) = \sum\nolimits_{\eta\mu \in \cZ_2^n}\:\: \e^{-C_{\eta\mu }
}\:P_\eta\: \rho(0)\: P_\mu\:, 
\een
where $P_\eta$ for $\eta \in \cZ_2^n \equiv \{0,1\}^n \equiv \{ \uparrow, \downarrow\}^n$ is
the projector on the state $\sket{\eta} = \sket{\eta_0}\dots
\sket{\eta_{n-1}}$, and $C_{\eta\mu}$ are real \cite{comment}
coefficients given by
\be
C_{\eta\mu}  =  \sum\nolimits_{lm} (\eta_l-\mu_l)(\eta_{m} - \mu_{m})
\:\Gamma_{|\br_l-\br_{m}|}\:. 
\ee 
The distance-dependent decoherence-parameter \cite{to_be}
\ben\label{Gamma_r}
\Gamma_r = A\int_0^\infty \!\!\! d\omega \: \omega^s
\fr{1-\cos \omega \tau}{\omega^2} \coth\left( \fr{\omega}{2 T} \right) 
\fr{\sin \omega r}{\omega r} \:e^{-\omega/\Omega}
\een
is a positive, monotonously decaying function of $r$ if $s<2$. 
Up to an singularity at $r=\tau$, the same holds for  
$2 \le s < 3$ if temperature is high, $T\gg \tau\inv,
r\inv$ \cite{to_be}. 
We confine our considerations to $s < 3$
and note that in quantum optical settings typically $s=1$, the Ohmic
case, where $\Gamma_r(\tau)$ is linear in $\tau$ for large $\tau \gg 1/T$ \cite{books}.

To make the model manageable we simplify the coefficients $C_{\eta\mu}$
by setting all distances $|\br_l - \br_{m}|_{l\neq m}$ to the maximum distance $r$.
Keeping in mind that this simplification lowers the effects of
correlations, we obtain 
\ben\label{coeff}
C_{\eta\mu} = \hw{\eta \oplus \mu} (\Gamma_0 - \Gamma_r ) + 
\left(\hw{\eta} - \hw{\mu} \right)^2 \Gamma_r\:, 
\een
where $\hw{\eta}$ denotes the Hamming weight \cite{macwilliams_sloan} of $\eta$,
and $\oplus$ means a bitwise addition modulo 2 in $\cZ_2^n$.

Although representation Eq.~\Ref{projector_rho} looks simple, it  
turns out to be rather cumbersome for our further calculations. 
A much better one can be given in terms of operators $Z_\nu$
that are defined for $\nu =(\nu_0\dots\nu_{n-1}) \in \cZ_2^n$ as
product of exactly those $\sigma_{z,l}$ where $\nu_l=1$. By the
structure of the interaction 
Hamiltonian it is evident that then 
\ben\label{Z_rep}
\rho(\tau) = \sum\nolimits_{\nu, \nu' \in \cZ_2^n} \:\:\alpha_{\nu \nu'}(\tau) \:Z_\nu\: \rho(0)\:
Z_{\nu'}\:, 
\een
with suitable coefficients $\alpha_{\nu\nu'}(\tau)$. To determine them we
let $\rho(0)= \sket{X}\sbra{X}$ be the projector on the totally $x$
polarized spin state and compare the expressions obtained in both representations.
In this way we see that $\alpha_{\nu\nu'}$ and $\e^{-C_{\eta\mu}}$ are
related by a Fourier transformation in $\cZ^n_2\times \cZ^n_2$,
\ben
\alpha_{\nu \nu'} = \fr{1}{4^n} \sum\nolimits_{\eta\mu \in \cZ_2^n}\:\: (-1)^{\nu
  \cdot \eta + \nu' \cdot \mu} 
\:\e^{-C_{\eta \mu}}\ \label{ft}
\een 
($\nu \cdot \eta$ denotes the standard inner product in $\cZ_2^n$).
Later on we will only need the diagonal coefficients 
$\alpha_{\nu\nu}$, which solely depend on the Hamming weight
$\hw{\nu}$. For $C_{\eta\mu}$ according to Eq.~\Ref{coeff} the sum \Ref{ft}
can be easily carried out if we eliminate the squared term in the
exponent by the identity 
$ 
\e^{-(\hw{\eta}-\hw{\mu})^2 \Gamma} = \fr{1}{\sqrt{\pi \Gamma}} \int dx
\:\: \e^{-\fr{x^2}{\Gamma} + 2ix\hw{\eta}  -2ix\hw{\mu}}
$. After some algebra we arrive at 
\ben\label{int_rep}
\alpha_{\nu\nu} \equiv \beta_\hw{\nu} =  \int \fr{dx}{\sqrt{\pi
    \Gamma_r}} \:
\e^{-\fr{x^2}{\Gamma_r}}\:\: p_x^\hw{\nu} \:\:(1-p_x)^{n-\hw{\nu}}\:,
\een
where we introduced the $x$-dependent ``probability'' 
\ben\label{pofx}
p_x = (1- \e^{-\Gamma_0+\Gamma_r} \cos 2 x)/2 \:\in [0,1] \:.
\een
Thus, we have a convenient description of the
decoherence process at hand and can now turn to CSS codes.

The key idea of quantum error-correction is to encode the
information of $k$ logical qubits in an appropriate subspace $\cC$ of the
Hilbert space $\cH_n$ associated with $n>k$ physical qubits. $\cC
\subset \cH_n$ is called a quantum code of length $n$ and size
$k$. Error operations that 
unitarily map $\cC$ on different cosets of $\cC$ can then be detected
and corrected. A {\em CSS} quantum code is constructed on two 
linear classical codes \cite{macwilliams_sloan} (i.e.\ $\cZ_2$-linear spaces)  $C_2 
\subset C_1 \subset \cZ_2^n$ by 
\be
 \cC = \mbox{span}\{ \ket{Q} \}_{Q\in C_2^\perp/C_1^\perp} \subset \cH_n\:
\ee
where $C_i^\perp$ is the orthogonal space of $C_i$ with respect to the
inner product in $\cZ_2^n$, and the state vectors $\ket{Q}$ are
\be
\ket{Q}= \fr{1}{\sqrt{ |C_1|}} \sum\nolimits_{y\in C_1}\:\: Z_q \sket{y}\:, \quad 
q\in Q \:.
\ee 
$\cC$ encodes $k= \log_2 |C_2^\perp/C_1^\perp| = \dim C_1 - \dim C_2$
logical qubits in $n$ physical qubits
\cite{calderbank_shor_steane,qec_general}. The error correcting
capability of $\cC$ is determined by the minimum weights
\cite{macwilliams_sloan} $d_1$ and
$d_1^\perp$ of $C_1$ resp.\ $C_1^\perp$. An error correction scheme
using $\cC$ can correct up to $t=[\fr{d-1}{2}]$ universal qubit
errors, where 
$d = \min\{d_1,d_1^\perp\} $. This characterizes $\cC$ as an $[n,k,d]$ code.
With $P$ being the projector on $\cC$, the error correcting operation
$\cR$ associated with $\cC$ reads
\ben\label{correction}
\cR(\rho) = \sum\nolimits_{\nu,\mu \in \cZ_2^n, \hw{\nu},\hw{\mu} \le t} \:\:P X_\mu
Z_\nu \rho Z_\nu X_\mu P\:
\een
($X_\mu$ is analogously defined as $Z_\nu$ with $\sigma_{x,l}$ instead
of $\sigma_{z,l}$). 

CSS codes have been used to demonstrate the existence of efficient
(``good'') quantum error correcting codes \cite{calderbank_shor_steane}, as specified by a
theorem of Calderbank and Shor:

{\em Theorem (Calderbank and Shor): For sufficiently large $n$ there
exists always an $[n,k,d]$ CSS code satisfying
\ben\label{css_rate}
k/n > 1 - 2 H_2 (d/n) \equiv R_{css}(d/n)\:.
\een}
($H_2(x) = -x \log_2 x - (1-x) \log_2(1-x)$ is the binary entropy function.)
Provided that the error-correcting operation $\cR$ for such codes is
error-free, this theorem can be rephrased in a more pragmatic fashion:
An $n$-qubit register that is perturbed in at most $t$ qubits
can be used to perfectly restore $k = [ n R_{css}(\fr{2t+1}{n})]$
logical qubits that have been encoded in an appropriate $[n,k,2t+1]$ CSS 
code. 

Of course, a real and noisy physical $n$-qubit register is unlikely to guarantee
the prerequisite of the theorem. It therefore does not necessarily
provide a practical solution of the 
decoherence problem. One has to demonstrate that also under more realistic
conditions the performance of the error-correcting code is still
sufficient. This has been shown in refs.\ \cite{calderbank_shor_steane} for the
independent error model. We will now investigate how CSS error
correction performs when the errors are correlated.

To this end we consider an arbitrary state $\sket{\Psi}_L =
\sum_{\eta} \psi_\eta \lket{\eta}$ of 
an abstract $k$-qubit register with $2^k$ orthonormal logical basis states
$\lket{\eta}$, $\eta\in \cZ_2^k$. The logical state $\sket{\Psi}_L$ is
encoded in an $[n,k,d]$ CSS code $\cC$ as a vector $\sket{\Psi} = \sum_\eta \psi_\eta
\sket{Q_\eta}\in \cC$ according to some bijective mapping  
$
 \lket{\eta} \mapsto \sket{Q_\eta}\:.
$
The encoded state $\rho_\Psi = \sket{\Psi}\sbra{\Psi} $ is first subjected
to the noise operation defined by Eq. \Ref{Z_rep} and then corrected
by $\cR$. This results in a final state $\rho'_\Psi = \cR   
(\rho_\psi(\tau)) $ which we compare with the original state $\rho_\Psi$ by
the fidelity $F(\rho_\Psi,\rho'_\Psi)$, which here is
\ben\label{fidelity}
\sbrackets{\Psi| \rho'_\Psi | \Psi} 
= \sum_{ \hw{\mu} \le t} \sum_{\nu\nu'} \alpha_{\nu\nu'}
\sbrackets{\Psi| Z_{\mu+\nu} | \Psi} \sbrackets{\Psi| Z_{\mu+\nu'} |\Psi}\:. 
\een
Since the square root of $1-F$ defines a proper distance measure for density
matrices \cite{gilchrist}, we call
\ben\label{delta_definition}
\Delta_\Psi \equiv 1 - F(\rho_\Psi, \rho'_\Psi)\:
\een
the residual error after CSS error correction.

Here we are not interested in specific codes but in the general
properties of CSS codes. We therefore continue by taking the average 
over practically all good CSS codes of a given length $n$ and a size $k$. This
is technically possible because of the following  

{\em Theorem: For any positive $\eps$ the $[n,k,d]$ CSS code
associated with randomly chosen classical codes $C_2 \subset C_1
\subset \cZ_2^n$ with $\dim C_2 = [\fr{n-k}{2}]$ and $\dim C_1 =
[\fr{n+k}{2}]$ satisfies 
\be
k/n \ge (1-\eps) R_{css}(d/n)\:
\ee
with a probability larger than 
$
1 - 2^{-n \left(\alpha \eps + \cO(n\inv) \right)}\:,
$
where $\alpha$ is  a positive constant independent of $n$ and $k$. }

The situation is thus very much like the one in classical coding
theory: randomly chosen 
subspaces of $\cZ_2^n$ yield codes that asymptotically reach the
Gilbert-Varshamov bound \cite{macwilliams_sloan}. The theorem can be
proven \cite{to_be} along the same lines as the proof in
\cite{coffey} for the classical case, plus application of MacWilliams'
theorem \cite{macwilliams_sloan}.

Let an average over all CSS codes of length $n$ and   
size $k$ be defined via the uniform average over all pairs
$C_2 \subset C_1$ of linear subspaces in $\cZ_2^n$ with dimensions 
$\dim C_{2/1} = [\fr{n \mp k}{2}]$. According to the theorem, this average
can be understood as an average over essentially all CSS codes of length $n$ that
asymptotically correct up to $t$  qubit errors, where $t =
[\fr{d-1}{2}]$ is determined by $k/n = R_{css}(d/n)$. 
We denote this average by $[\dots ]^{n}_{t}$.

Applying the average to Eq. \Ref{fidelity} leads us finally  to
\ben\label{delta}
[\Delta_\Psi]^{n}_{t} = \sum\nolimits_{\hw{\nu} > t} \:\alpha_{\nu\nu} = \sum\nolimits_{w=t+1}^n
\:\binomial{n}{w} \beta_w\:, 
\een
where we used the completeness relation $\sum_\nu \alpha_{\nu\nu} = 1$, and
suppressed terms of order $2^{-|\cO(n)|}$.   
The physical interpretation of this expression is that more than $t$
simultaneous qubit errors cannot be corrected, and therefore
coefficients $\alpha_{\nu\nu}$ with $\hw{\nu}> t$ contribute to the residual error.
Terms with non-diagonal coefficients $\alpha_{\nu\neq
  \nu'}$ turn out to be suppressed by a factor $2^{-{\fr{n+k}{2}}}$ 
and therefore do not significantly contribute. Up to these
exponentially small corrections the code averaged residual error is
independent of the encoded logical state $\lket{\Psi}$. 

In the following we consider $[\Delta]^{n}_{t}$ for a fixed ratio $ q
\equiv \fr{t+1}{n}$.
Inserting Eq.\ \Ref{int_rep} into \Ref{delta} immediately leads to
\ben\label{general_delta}
[\Delta]^{n}_{q n} = 
\int \!\!\fr{dx}{\sqrt{\pi \Gamma_r}}\:   \e^{-\fr{x^2}{\Gamma_r}} 
\sum\nolimits_{w=qn}^{n}
\binomial{n}{w}p_x^w (1-p_x)^{n-w}\:.
\een
From this general formula for the residual error we can now draw 
conclusions on the performance of CSS codes. 

Independent errors correspond to the case of diverging distance 
$r$, where the decoherence parameter $\Gamma_r$ vanishes.
In this limit the Gaussian in \eref{general_delta} shrinks to a
normalized delta peak at $x=0$. Hence, 
\ben\label{delta_independent}
[\Delta]^{n}_{q n} = \sum\nolimits_{w=q n}^n \binomial{n}{w}
p_o^w (1-p_o)^{n-w}\:,
\een
 where
$p_o=(1-\e^{-\Gamma_o})/2$ is the error probability of a single
spin.
This expression, formerly derived by Calderbank and Shor
\cite{calderbank_shor_steane}, implies 
a binomially distributed residual error that decays exponentially with
$n$ as long as $p_o<q$.
The point is that the constraint on the single-spin
decoherence $\Gamma_0 \approx 2 p_o < 2 q $ is
$n$-independent. In this sense,  CSS codes provide 
scalable quantum-error correction with an exponentially small residual
error \cite{calderbank_shor_steane}.  

For finite $\Gamma_r$ the asymptotic value of the residual error can
be easily derived from \eref{general_delta} by the 
observation that in the limit $n\to\infty$ 
the sum over $w$ vanishes if $p_x<q$ and equals
unity if $p_x>q$. We obtain
\ben\label{delta_erfc}
\lim_{n\to\infty}[\Delta]^{n}_{qn} = \int_{p_x > q}\fr{dx}{\sqrt{\pi \Gamma_r}}\:
\e^{-\fr{x^2}{\Gamma_r}}  \:\:\sim \:\: \mbox{erfc}\sqrt{ \fr{q}{\Gamma_r}
}\:,
\een
where the last approximation is good for $\Gamma_r, \Gamma_0 \ll q $. 
In sharp contrast to the independent errors discussed above 
for any finite $\Gamma_r$ (i.e., for any finite $r$) the residual error
converges for large $n$ to a finite constant (cf.~Fig.~\ref{fig1}).

According to quantum complexity theory \cite{bernstein_vazirani} the 
total error in an $m$ step quantum computation must be less than
$m^{-|\cO(1)|}$ in order to produce useful results. If we therefore demand
that the residual error must be limited by a $\Delta_{max}(n) \sim b/n^\mu$ with some
positive $b$ and $\mu$, we deduce from
\Ref{delta_erfc} that $\Gamma_r$ must obey
\ben\label{condition}
\Gamma_r < \Gamma_{max}(n) \sim  q/(\mbox{const} + \mu \ln n)\:.
\een
This is no longer independent of $n$ as for the uncorrelated model
but approaches zero in the limit $n\to\infty$, although slowly with
the inverse logarithm of $n$.  

A further discussion must take into account that also
$\Gamma_r$ by \eref{Gamma_r} depends on $n$ via the
$n$-dependence of the maximum distance $r$ and the observation time
$\tau$. It is plausible to assume $r \propto n^y$ with some exponent
$y \ge 1/3$, and also $\tau \propto n^y $,  
because the observation time $\tau$ must scale at least linearly with
the time needed to transmit signals between qubits. 
We therefore set $r = a r_0$, $\tau = a \tau_0$, where $a=(n/n_0)^y$,
and write
\be
\Gamma_r \equiv \Gamma(s, a r_0, a \tau_0, T, \Omega)= a^{1-s} \Gamma(s,r_0,\tau_0,
aT, a\Omega)\:.
\ee
The second equality is obtained by rescaling the integral \Ref{Gamma_r}.
Inspection of Eq.\ \Ref{Gamma_r} also reveals that 
for large effective temperatures $aT \gg r_0\inv,
\tau_0\inv$ the decoherence parameter is linear in temperature, and
virtually independent on $\Omega$. For large $n$ condition \Ref{condition}
is therefore equivalent to
\ben\label{inequality} 
\Gamma(s,r_0,\tau_0,T) < \mbox{$ \fr{q}{ \mbox{{\small const}} + \mu\ln n}$}\:
(\mbox{$\fr{n}{n_0}$})^{y(s-2)}\:. 
\een
The inequality is violated if $s\le 2$ and $n$ is large. Consequently,
in this regime the residual error exceeds $\Delta_{max}(n)$.
On the other hand, for $s>2$  conditions \Ref{inequality} and 
\Ref{condition} are satisfied for large $n$, meaning 
that here a sufficiently small residual error can be attained.
We conclude that for interaction paramter $s > 2$ scalable
CSS error-correction is possible, while for $s \le 2$ it is not.  

Apparently, noise correlations introduced by the spin-boson
interaction are a strong hindrance for small values of $s$, whereas 
they are less harmful for larger $s$. 
An intuitive explanation is that with increasing $s$ spectral weight in
$J(\omega)$ is shifted from lower to higher frequencies, and therefore
the larger the value of $s$, the faster the disturbing noise
correlations decay with distance. 

Note that the physical model we use is lacking dissipative couplings,
which corresponds to a systematic underestimation of quantum
noise. So, for a spin-boson model including dissipative couplings
the residual error will be larger than the one calculated here. 
A second caveat is that we considered perfect quantum
error-correction, while real systems will have to rely on  
{\em fault-tolerant} error-correcting schemes. 
It is not clear to what extend the results presented here do apply to
those schemes. However, a naive guess might
be based on the observation that fault-tolerant quantum computation
is, in some sense, simulating 
perfect quantum error-correction with imperfect quantum gates. This
suggests that fault-tolerant schemes cannot better perform than
perfect error-correction investigated here. Following these arguments,
it is important to note that our results are not at odds with a
recently proven threshold theorem \cite{aliferis}. Here the spin-boson 
interaction is unbounded \cite{terhal_burkard} and therefore does not
belong to the class of interactions considered in Ref.\ \cite{aliferis}.

We wish to thank H.~Moraal for numerous valuable discussions and
M.~R.~Zirnbauer for critical reading of the manuscript. The work is
supported by the SFB TR/12.


\begin{thebibliography}{10}
\bibitem{units} We use units in which $\hbar=1,  k_B=1$ and $ c = \omega_k/|k| = 1$.
\end{thebibliography}
\end{document}